\documentclass[aps,prl,twocolumn,superscriptaddress,showpacs]{revtex4}
\usepackage{amsmath,amssymb}
\usepackage{graphicx}
\usepackage{bm}

\newcommand{\pdag}{\phantom{\dag}}
\newcommand{\veps}{\varepsilon}
\newcommand{\rhoeff}{\rho_{\mathrm{eff}}}
\newcommand{\chiimp}{\chi_{\mathrm{imp}}}

\begin{document}

\title{Tunable pseudogap Kondo effect and quantum phase transitions in
Aharonov-Bohm interferometers}

\author{Luis G.\ G.\ V.\ Dias da Silva}
\email{diasdasilval@ornl.gov} \affiliation{Materials Science and
Technology Division, Oak Ridge National Laboratory, Oak Ridge,
Tennessee, 37831, USA and Department of Physics and Astronomy,
University of Tennessee, Knoxville, Tennessee 37996, USA}

\author{Nancy Sandler}
\affiliation{Department of Physics and Astronomy, Nanoscale and
Quantum Phenomena Institute, Ohio University, Athens, Ohio
45701-2979, USA}

\author{Pascal Simon}
\affiliation{Laboratoire de Physique et Mod\'elisation des Milieux
Condens\'es, CNRS et Universit\'e Joseph Fourier, 38042 Grenoble,
France} \affiliation{Laboratoire de Physique des Solides, CNRS
UMR-8502, Universit\'e Paris Sud, 91405 Orsay Cedex, France}

\author{Kevin Ingersent}
\affiliation{Department of Physics, University of Florida, P.O.\
Box 118440, Gainesville, Florida 32611-8440, USA}

\author{Sergio E.\ Ulloa}
\affiliation{Department of Physics and Astronomy, Nanoscale and
Quantum Phenomena Institute, Ohio University, Athens, Ohio
45701-2979, USA}

\date{\today}

\begin{abstract}
We study two quantum dots embedded in the arms of an Aharonov-Bohm
ring threaded by a magnetic flux. The system can be described by
an effective one-impurity Anderson model with an energy- and
flux-dependent density of states. For specific values of the flux,
this density of states vanishes at the Fermi energy, yielding a
controlled realization of the pseudogap Kondo effect. The
conductance and transmission phase shifts reflect a nontrivial
interplay between wave interference and interactions, providing
clear signatures of quantum phase transitions between Kondo and
non-Kondo ground states.
\end{abstract}

\maketitle

Nanoscale quantum-dot devices are a formidable tool for probing
the inherent quantum-mechanical nature of electrons.
Manifestations of quantum electronic properties in these devices
include wave interference in Aharonov-Bohm (AB) rings
\cite{QD:interference,Aharony:156802:2003,Zaffalon:226601:2008}
and many-body phenomena such as the Kondo effect (the screening of
a localized magnetic moment by conduction electrons)
\cite{QD:Kondo,QD:Kondo-rings,QD:QPTs,Zaffalon:226601:2008} and
quantum phase transitions (QPTs) \cite{QD:QPTs}. The interplay
between quantum interference and the Kondo effect can be studied
by inserting a quantum dot in an AB ring, as shown both
experimentally \cite{QD:Kondo-rings} and theoretically
\cite{QD:Kondo-rings-theory,Lopez:115312:2005,Simon:245313:2005}.

This Letter focuses on a system in which two quantum dots are
embedded in the same AB ring. Interesting effects have been
proposed \cite{Lopez:115312:2005} in cases where both dots are in
the Kondo regime. Here, we consider instead a device in which the
presence of one, effectively noninteracting dot creates for a
second, Kondo-regime dot, an energy-dependent effective density of
states that depends on the magnetic flux applied through the ring.
Varying this flux can dramatically affect the Kondo state in the
interacting dot, causing the Kondo temperature $T_{K}$---the
characteristic energy scale of the Kondo state---to range over
many orders of magnitude.

This two-dot AB device can also realize the conditions necessary
for observation of the \textit{pseudogap Kondo effect}
\cite{Withoff:1835:1990,pseudogap-Anderson}, in which coupling of
a magnetic impurity to a power-law-vanishing density of conduction
states gives rise to a pair of QPTs between Kondo ($T_K>0$) and
non-Kondo ($T_K=0$) phases. Pseudogap Kondo physics has previously
been predicted to occur in double-quantum-dot devices
\cite{Dias:096603:2006,Dias:0804.0805:2008}, but the ring geometry
of the present setup allows a deeper exploration of the interplay
between coherent quantum interference and the Kondo effect. The
conductance and transmission phase shift through the system
exhibit clear signatures of each zero-temperature transition
within a quantum-critical region that extends up to temperatures
of order the maximum Kondo scale of the interacting dot. This
robustness plus the relative ease of experimental control make the
proposed device very promising for experimental investigation of
pseudogap Kondo physics.

\begin{figure}[t]
\begin{minipage}[c]{0.48\columnwidth}
\includegraphics*[width=\columnwidth]{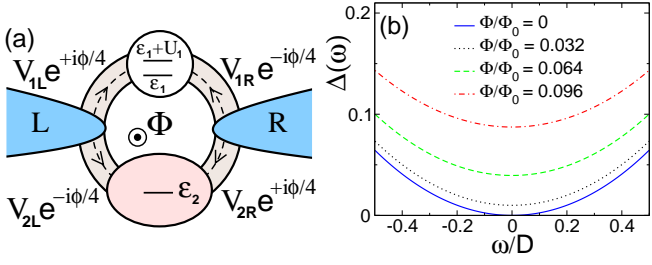}
\end{minipage}
\begin{minipage}[c]{0.5\columnwidth}
\includegraphics*[width=\columnwidth]{fig1b_selfenergy1.eps}
\end{minipage}
\caption{\label{fig:DDot_flux}%
(Color online)
(a) Schematic of the AB interferometer with embedded quantum dots, coupled
to external leads ($L$ and $R$) and threaded by a magnetic flux $\Phi$.
(b) Hybridization function $\Delta(\omega)$ for $\veps_2=0$ and different
values of $\Phi$.}
\end{figure}

\textit{Model.}---Quantum dots (``1'' and ``2'') are embedded in opposite
arms of an AB interferometer that is connected to left (``$L$'') and right
(``$R$'') metallic leads, as shown in Fig.\ \ref{fig:DDot_flux}(a).
Dot 1 is in a Coulomb blockade valley and is occupied by an odd number of
electrons, while dot 2 has a single noninteracting level in resonance with
the leads. An external AB flux $\Phi$ passes through the interferometer,
causing a phase difference $\phi=2\pi \Phi/\Phi_0$ ($\Phi_0=hc/e$) between
electrons that tunnel from $L$ to $R$ via dot 1 and those that tunnel via dot 2.
Provided that the flux through each quantum dot (as opposed to the entire ring)
is much smaller than $\Phi_0$, orbital effects can be neglected. The low
$g$-factor in typical GaAs devices allows one also to disregard the Zeeman
splitting in the dots. In this approximation, the Hamiltonian for the setup is
\begin{multline}
\label{Eq:H}
H = \sum_{j,\sigma} \veps_j a^{\dag}_{j\sigma} a^{\pdag}_{j\sigma}
  + U_1 a^{\dag}_{1 \uparrow} a^{\pdag}_{1 \uparrow}
        a^{\dag}_{1 \downarrow} a^{\pdag}_{1 \downarrow}
  + \sum_{\ell,\mathbf{k},\sigma} \! \veps_{\ell\,\mathbf{k}}
    c^{\dag}_{\ell\mathbf{k}\sigma} c^{\pdag}_{\ell\mathbf{k}\sigma}
    \\[-.5ex]
  + \sum_{j,\ell,\mathbf{k},\sigma} \bigl( W_{j\ell} \, a^{\dag}_{j\sigma}
    c^{\pdag}_{\ell\mathbf{k}\sigma}+ \text{H.c.}\bigr) \, ,
\end{multline}
where $a_{j\sigma}$ destroys a spin-$\sigma$ electron in dot $j$
($j=1,2$) and $c_{\ell\,\mathbf{k}\sigma}$ destroys a
spin-$\sigma$ electron of wave vector $\mathbf{k}$ and energy
$\veps_{\ell\,\mathbf{k}}$ in lead $\ell$ ($\ell=L,R$). Each lead
is assumed to have a constant density of states
$\rho(\veps)=\rho_0\Theta(D-|\veps|)$, as well as a local
($\mathbf{k}$-independent) coupling to the dots. The gauge degree
of freedom allows one to write $W_{1 L} = V_{1 L} e^{+i\phi/4}$,
$W_{1 R} = V_{1 R} e^{-i\phi/4}$, $W_{2 L} = V_{2 L}
e^{-i\phi/4}$, and $W_{2 R} = V_{2 R} e^{+i\phi/4}$ where
$V_{j\ell}$ is real. For simplicity, we consider symmetric
couplings $V_{jR}=V_{jL}\equiv V_j/\sqrt{2}$.

At small bias and low temperatures, transmission through an interacting
system can be described by a
Landauer-like formula \cite{Meir:2512:1992}. The conductance $g$ and the
transmission phase shift $\theta_t$ of the device
are given by
\begin{align}
\label{Eq:g}
g& = \frac{2e^2}{h} \int d\omega \,
     \left(-\frac{\partial f}{\partial\omega}\right) \,
     |t_{LR}(\omega)|^2 \, , \\
\label{Eq:theta}
\theta_t& = \arg \, \int d\omega  \,
     \left(-\frac{\partial f}{\partial\omega}\right) \,
     t_{LR}(\omega) \, ,
\end{align}
where $f(\omega,T)$ is the Fermi function at energy $\omega$
(measured from the Fermi level) and temperature $T$, and
$t_{LR}(\omega)\!=\!2\pi\rho_0 \sum_{ij} W^{*}_{iL} G_{ij}(\omega)
W_{jR}$ is the transmission coefficient. Here, $G_{ij}(\omega) =
-i\int_0^{\infty} \! dt \,  e^{i\omega t} \langle \{ a^{\pdag}_{i
\sigma}(t), a^{\dag}_{j \sigma}(0)\}\rangle$ is a standard
retarded Green's function.

The dot-1 Green's function (calculated in the presence of dot 2 and the leads,
and taking the $U_1$ interaction into full account) can formally be written
$G_{11}(\omega) = [\omega - \veps_1 - \Sigma_{11}^*(\omega) -
\Sigma^{(0)}_{11}(\omega)]^{-1}$, where $\Sigma_{11}^*$ and $\Sigma_{11}^{(0)}$
are, respectively, the interacting and noninteracting contributions to the
self-energy. Standard equations of motion techniques can be used to express
the remaining $G_{ij}$'s in terms of $G_{11}$ and known quantities, and to
obtain the exact result
\begin{equation}
\label{Eq:Sigma0}
\Sigma^{(0)}_{11}
= \sum_{\ell,\mathbf{k}}
  \frac{|W_{1\ell}|^2}{\omega \! - \! \veps_{\ell\mathbf{k}}} + \!\!\!
  \sum_{\ell,\ell',\mathbf{k},\mathbf{k}'} \!\!
  \frac{W^{\phantom{*}}_{1\ell}W^*_{2 \ell}}{\omega \!-\!\veps_{\ell\mathbf{k}}}
  \frac{1}{{\omega \! - \! \veps_2 \! + \! i\Delta_2}}
  \frac{W^{\phantom{*}}_{2 \ell^\prime} W^*_{1 \ell^\prime}}{\omega \! - \!
    \veps_{\ell^\prime\mathbf{k}^\prime}},
\end{equation}
where $\Delta_j = \pi \rho_0 V_j^2$. The first term in Eq.\
\eqref{Eq:Sigma0} describes the effect on dot 1 of coupling purely
to the leads, while the second term represents an indirect
coupling between the dots.
In the wide-band limit $|\omega|\ll D$, these processes combine to
yield an energy-dependent hybridization width
$-\mbox{Im}\Sigma^{(0)}_{11}(\omega) \equiv \pi\rhoeff(\omega)
V_1^2$, with
\begin{equation}
\label{Eq:rhoeff}
\rhoeff(\omega)=\rho_0 \frac{(\omega - \veps_2)^2 + \Delta_2^2
  \sin^2(\pi\Phi/\Phi_0)}{(\omega-\veps_2)^2 + \Delta_2^2} .
\end{equation}
Then $G_{11}(\omega)$ corresponds to the Green's function of a
single Anderson impurity coupled to a density of conduction states
$\rhoeff(\omega)$ that is periodic in the applied flux. Note that
$\rhoeff(\omega)=\rho_0$ for $\Phi=(n+\frac{1}{2})\Phi_0$, where
$n$ is any integer. More generally, $\rhoeff(\omega)\simeq\rho_0$
for $|\omega-\veps_2|\gg\Delta_2$, dipping to
$\rhoeff(\omega)\simeq \rho_0\sin^2(\pi\Phi/\Phi_0)$ for
$|\omega-\veps_2|\ll\Delta_2$. For special cases where $\veps_2=0$
and $\Phi=n\Phi_0$, $\rhoeff(\omega)$ vanishes at the Fermi energy
as $\omega^2$ [solid line in Fig.\ \ref{fig:DDot_flux}(b)], and
the low-energy physics is that of the pseudogap Anderson model
\cite{pseudogap-Anderson}. In all other cases, $\rhoeff$ is
metallic and one recovers a conventional Anderson model, albeit
one with a field-modulated impurity-host coupling.

This analysis raises the intriguing prospect of realizing a \textit{flux-tuned}
pseudogap in a two-dot AB ring device. We have solved the effective
one-impurity model suggested by Eq.\ \eqref{Eq:rhoeff} using the numerical
renormalization-group method \cite{KrishnamurthyWW80_1,Bulla:395:2008} to
obtain properties of the full system.
Below, we fix $U_1=0.5D$, $\Delta_1=0.05D$, and $\Delta_2=0.02D$, and show
results for different values of $\veps_1$ and $\veps_2$ (controlled in
experiments by plunger gate voltages on dots 1 and 2, respectively) and
of the AB flux $\Phi$.

\begin{figure}[tbp]
\includegraphics*[width=0.9\columnwidth]{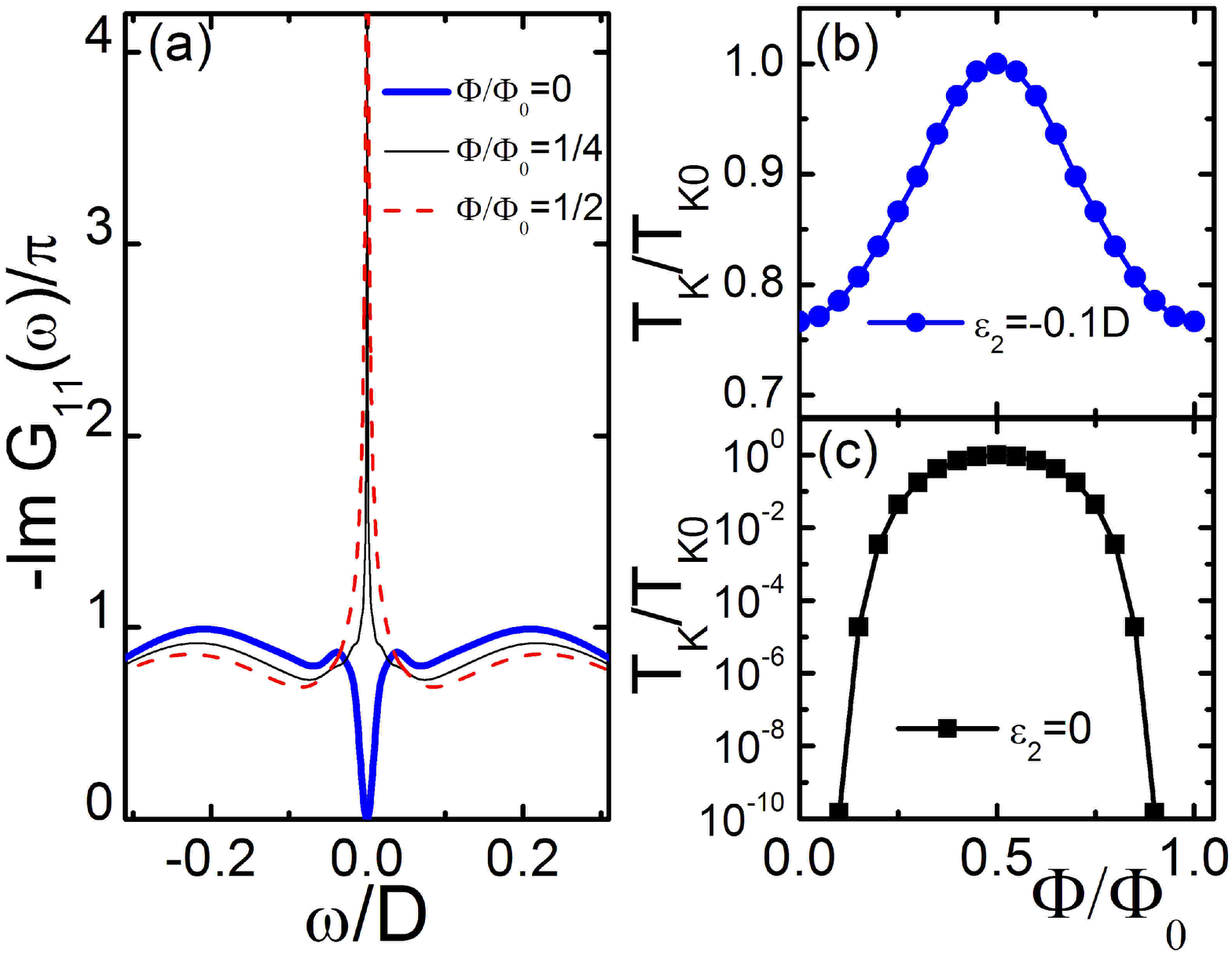}
\caption{\label{fig:G_phi}%
(Color online) (a) Dot-1 spectral function $A_{11}(\omega)$ for
$\veps_1=-U_1/2$, $\veps_2=0$, and different magnetic fluxes
$\Phi$. (b,c) Kondo temperature $T_K/T_{K0}$ vs $\Phi/\Phi_0$ for
$\veps_1=-\frac{1}{2}U_1$ and (b) $\veps_2=-0.1D$, (c)
$\veps_2=0$. The characteristic many-body scale
$T_{K0}=T_K(\veps_1=-\frac{1}{2}U_1,\Phi=\frac{1}{2}\Phi_0)$ is
independent of $\veps_2$.} \vspace{-1em}
\end{figure}

\textit{Variation of the Kondo scale.}---Figure \ref{fig:G_phi}(a) shows the
dot-1 spectral density $A_{11}(\omega) = -\pi^{-1} \mathrm{Im}G_{11}(\omega)$
for several $\Phi$ values at the special point $\veps_1=-U_1/2$, $\veps_2=0$
where the system exhibits strict particle-hole ($p$-$h$) symmetry. For a general
flux, $A_{11}(\omega)$ features a Kondo resonance centered on $\omega = 0$.
For $\Phi= n\Phi_0$, however, $A_{11}(\omega)$ vanishes at $\omega=0$, signaling
suppression of the Kondo effect by the pseudogap in $\rhoeff(\omega)$
\cite{Dias:096603:2006}.

The Kondo resonance width is proportional to the Kondo temperature
$T_K$, which we define in terms of the impurity susceptibility via
the condition $T_K\chiimp(T_K)=0.0701$ \cite{KrishnamurthyWW80_1}.
$T_K$ values varying over three orders of magnitude under an
applied magnetic field have
been predicted for small AB
rings containing one
quantum dot \cite{Simon:245313:2005}.
The present setup can greatly amplify this variation.
For $|\veps_2|\gtrsim \Delta_2$
[see, \textit{e.g.}, Fig.\ \ref{fig:G_phi}(b)],
the dip in $\rhoeff(\omega)$
around $\omega=\veps_2$ produces only a weak field-modulation of
$T_K$.
The range of $T_K$ is much greater for $|\veps_2|\lesssim\Delta_2$.
In the extreme case $\veps_2 = 0$
[Fig.\ \ref{fig:G_phi}(c)], $T_K$ varies
from $T_{K0}$
for $\Phi=(n\!+\!\frac{1}{2})\Phi_0$ to zero for
the pseudogap case $\Phi = n\Phi_0$.
Here and below,
$T_{K0}=T_K(\veps_1\!=\!-\frac{1}{2}U_1,\Phi\!=\!\frac{1}{2}\Phi_0)
\simeq 7\!\times\! 10^{-4}D$
is a characteristic Kondo scale for dot 1 in the absence of dot 2.

\begin{figure}[tbp]
\includegraphics*[width=0.9\columnwidth]{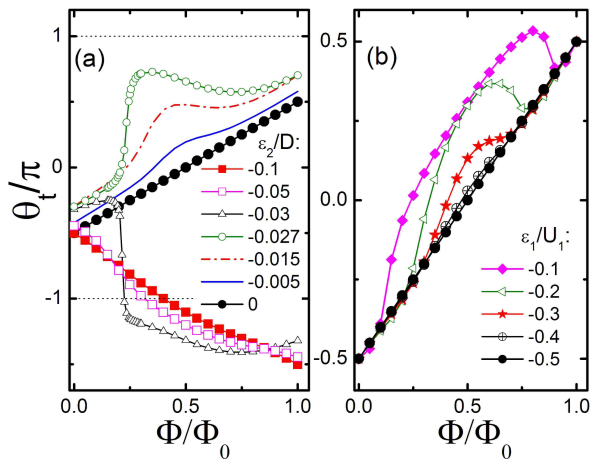}
\caption{\label{fig:theta_vs_Phi}%
(Color online) Phase shift $\theta_t$ vs AB flux $\Phi$ at $T=0.59
T_{K0}$ for (a) $\veps_1=-U_1/2$ and different values of $\veps_2$
and (b) $\veps_2 = 0$ and different values of $\veps_1$. In (a),
phase lapses $\Delta\theta_t\simeq \pm\pi$ occur around
$\veps^{\mathrm{pl}}_2 = -0.0292D$.}
\vspace{-1em}
\end{figure}

\textit{Quantum phase transitions.}---As noted in the
introduction, the presence of a pseudogap in $\rhoeff(\omega)$
gives rise to a pair of QPTs separating Kondo and local-moment
phases \cite{Dias:096603:2006,Dias:0804.0805:2008}. These QPTs
occur in the double-dot AB setup for $\Phi = n\Phi_0$ and $\veps_2
= 0$ when $\veps_1$ is tuned to one of two critical values
$\veps_{1c}^{\pm}$. The paragraphs below describe how the system
can be brought into the vicinity of one of these zero-temperature
transitions by measuring the transmission phase shift
$\theta_t(\Phi)$ and/or the conductance $g(\Phi)$ at relatively
high temperatures of order $T_{K0}$.

The first step in reaching the QPT is to bring the dot-2 level
$\veps_2$ to the Fermi energy. We find that this can be most
efficiently accomplished by monitoring $\theta_t(\Phi)$. Figure
\ref{fig:theta_vs_Phi}(a) plots $\theta_t$ at $T=0.59 T_{K0}$ over
the range $0\le\Phi\le\Phi_0$ for $\veps_1=-U_1/2$ and various
values of $\veps_2$. The most striking feature is the linear
variation of $\theta_t$ with $\Phi$ that can be used to identify
the target case $\veps_2=0$. The origin of this linearity can be
seen most readily at $T=0$, where for $\veps_2=0$,
$\theta_{t}=\pi(\Phi/\Phi_0-\frac{1}{2}) +
\bar{\theta}_t$, with
\begin{equation}
\label{Eq:barTheta}
\bar{\theta}_t = \tan^{-1}
 \frac{\Delta_1 \text{Re} \, G_{11}(0) \sin^2(\pi\Phi/\Phi_0)}
      {\Delta_1 \text{Im} \, G_{11}(0) \sin^2(\pi\Phi/\Phi_0)-1} .
\end{equation}
At the pseudogap points $\Phi=n\Phi_0$, $\sin(\pi\Phi/\Phi_0)=0$ and
$\bar{\theta}_t = 0$. Everywhere else, a conventional Kondo ground state
forms. The special case $\veps_1=-U_1/2$ and $\veps_2 = 0$ shown in
Fig.\ \ref{fig:theta_vs_Phi}(a) exhibits an exact $p$-$h$ symmetry that
ensures $\text{Re}\,G_{11}(0)=0$ and $\bar{\theta}_t = 0$ for all $\Phi$.

Figure \ref{fig:theta_vs_Phi}(a) also reveals interesting features
away from $\veps_2 = 0$.  For large $|\veps_2|$,
$\theta_t$ evolves with increasing $\Phi$ to pass through $-\pi$ from above;
since phase shifts
$\theta_t$ and $\theta_t\pm 2\pi$ are equivalent, any such curve
can instead be plotted with a phase jump from $-\pi$ to $\pi$, so
that in all cases, $\theta_t(\Phi_0)=\theta_t(0)+\pi$.
Around $\veps_2=\veps^{\mathrm{pl}}_2$, ``phase lapses''
$\Delta \theta_t \simeq \pm\pi$ (\textit{not} $\pm 2\pi$)
appear over narrow ranges of $\Phi$ \cite{phase_lapses}.
For $\veps_2>\veps^{\mathrm{pl}}_2$, $\theta_t$ does not pass through
$\pm\pi$, but rather varies smoothly between $\theta_t(0)$ and
$\theta_t(\Phi_0)=\theta_t(0)+\pi$.

For general $\veps_1$, $\veps_2$, and $T$, $\bar{\theta}_t \equiv
\theta_t - \pi(\Phi/\Phi_0 - \frac{1}{2})$ is small whenever
$T_K\ll T$, and is appreciably nonzero for $T_K\gtrsim T$. This is
illustrated in Fig.\ \ref{fig:theta_vs_Phi}(b), which plots the
phase shift at $T= 0.59 T_{K0}$ for $\veps_2 = 0$ and different
values of $\veps_1$. In each case, the Kondo temperature vanishes
for $\Phi=n\Phi$ and reaches its maximum value $T_{K,\max}$ at
$\Phi=(n+\frac{1}{2})\Phi_0$. With increasing $p$-$h$ asymmetry
(increasing $|\veps_1+U_1/2|$), $T_{K,\max}$ decreases and the
points of first noticeable deviation from linearity in $\theta_t$
vs $\Phi$ move closer to $\Phi=n\Phi_0$.

These results suggest an experimental procedure for tuning to the
pseudogap: Measure $\theta_t$ vs $\Phi$ for different dot-2
plunger gate voltages, holding all other parameters constant, and
seek to maximize the range of fluxes around $\Phi = n\Phi_0$ over
which the phase shift satisfies $\bar{\theta}_t = 0$. If one has
truly found the dot-2 gate voltage corresponding to $\veps_2 = 0$,
it should in general be possible to increase the flux range over
which $\bar{\theta}_t=0$ by stepping the plunger gate voltage on
dot 1 until one achieves $\veps_1\simeq -U_1/2$.

\begin{figure}[t]
\includegraphics*[width=0.95\columnwidth]{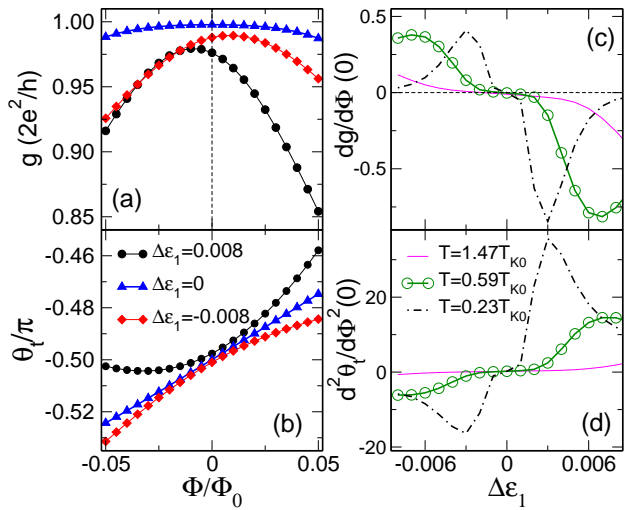}
\caption{\label{fig:pseudogap_transport}%
(Color online) Variation with $\Phi/\Phi_0$ of (a) the conductance
$g$ and (b) the phase shift $\theta_t$, for $\veps_2=0$, $T=0.59
T_{K0}$, and different $\Delta \veps_1 \equiv
\veps_1-\veps_{1c}^+$. For $\veps_2=0$, $\Phi = 0$, and different
temperatures $T$, both (c) $dg/d\Phi$ and (d) $d^2
\theta_t/d\Phi^2$ change sign at the critical value
$\Delta\veps_1=0$.}
\vspace{-1.5em}
\end{figure}

Once the dot-2 level is locked at the Fermi level, the system can be steered
through (or, at any $T>0$, above) a QPT by further fine-tuning of $\veps_1$,
guided by measurements of $g(\Phi)$ and $\theta_t(\Phi)$. We focus on the QPT
at $\veps_1=\veps_{1c}^+$, where $-U_1/2 < \veps_{1c}^+ < 0$, and define
$\Delta\veps_1 = \veps_1 - \veps_{1c}^+$. (A $p$-$h$ transformation maps the
system from $\veps_{1c}^+$ to the other QPT at $\veps_{1c}^-=-U_1-\veps_{1c}^+$.)
As illustrated in Fig.\ \ref{fig:pseudogap_transport}, the properties at
temperatures of order $T_{K0}$ reveal clear signatures of the $T=0$ transition
between the local-moment ($\Delta \veps_1<0$ and $\Phi=n\Phi_0$) and Kondo
($\Delta \veps_1>0$ and/or $\Phi\ne n\Phi_0$) phases.

At $\Delta \veps_1=0$ and
$\Phi=n\Phi_0$, the finite-temperature conductance reaches a
near-unitary value $g\simeq g_0$ [Fig.\
\ref{fig:pseudogap_transport}(a) for $n=0$]
while the transmission phase shift $\theta_t=-\pi/2$
[Fig.\ \ref{fig:pseudogap_transport}(b)]. However, these
characteristics may not be reliable experimental locators for the
underlying QPT because absolute measurements of $g$ or $\theta_t$
may be complicated by contributions from additional (spurious)
channels \cite{Aharony:156802:2003} or by the
presence of stray external flux that prevents accurate
identification of the point
$\Phi=n\Phi_0$.

The derivatives of the transport properties with respect to
applied flux provide a superior method for locating the
transition. The critical value $\Delta\veps_1 = 0$ is
distinguished by
two features around the pseudogap location $\Phi=n\Phi_0$: (i) $g$
is at a maximum [Fig.\ \ref{fig:pseudogap_transport}(a)] and (ii)
$\theta_t$ vs $\Phi$  is \textit{linear} over a significant window
in $\Phi$
with a temperature-dependent slope smaller than that of the line
$\bar{\theta}_t=0$
[Fig.\ \ref{fig:pseudogap_transport}(b)].
Figures\ \ref{fig:pseudogap_transport}(c) and
\ref{fig:pseudogap_transport}(d) show that at three different
temperatures of order $T_{K0}$, $dg/d\Phi|_{\Phi=0}$ and
$d^2\theta_t/d\Phi^2|_{\Phi=0}$ vs $\Delta\veps_1$ both pass
through zero at $\Delta \veps_1 =0$.

The most important conclusion to be drawn from Fig.\
\ref{fig:pseudogap_transport} is that features indicative of the
QPT are evident in the transport at least up to temperatures of
order $T_{K0}$, the characteristic scale of conventional Kondo
physics in the interacting dot, and one likely to be readily
accessible in experiments. Figures
\ref{fig:pseudogap_transport}(c) and
\ref{fig:pseudogap_transport}(d) also illustrate the general
property of continuous QPTs that with increasing temperature,
quantum-critical behavior extends over
a wider region of the
parameter space. The crossings of $dg/d\Phi|_{\Phi=0}$ and
$d^2\theta_t/d\Phi^2|_{\Phi=0}$ through zero
spread over a range of $\veps_1$
that grows roughly linearly with
$T$. Similar behavior (not shown) occurs
at small but nonzero $|\Phi|$ and/or $|\veps_2|$. Away from the
true critical values, however, the locations of key features (the
peak in $g$ and the sign change in $d^2\theta_t/d\Phi^2$)
depend on $T$, and below a crossover temperature
these features fade away
as the system enters the stable
Kondo or local-moment regime.

In summary,
we have studied the Kondo regime of two quantum
dots embedded in the arms of an Aharonov-Bohm ring threaded by a
magnetic flux. The system is described by an effective Anderson
model with an effective density of states that is modulated by the
external flux, allowing the Kondo temperature to be tuned over a
wide range. When the ring encloses an integer multiple of the
quantum of flux, the effective density of states vanishes at the
Fermi energy and the setup maps onto a pseudogap Anderson model.
The transmission phase shift at temperatures of order
the characteristic Kondo scale of a single, interacting dot
can be used to tune the
device to the pseudogap regime, where the phase shift and the
linear conductance exhibit clear finite-temperature signatures of
underlying zero-temperature phase transitions.

We acknowledge support under NSF-DMR grants 0312939 and 0710540
(University of Florida), 0336431, 0304314 and 0710581 (Ohio
University), and 0706020 (University of Tennessee/ORNL).

\end{document}